\documentclass[aps,showpacs,twocolumn,amsmath,amssymb,nofootinbib,prc]{revtex4-1}
\usepackage{epsfig}
\usepackage{multirow}
\usepackage{amsmath}
\usepackage{amssymb}
\usepackage{epstopdf}
\usepackage{color}
\usepackage{bm}
\usepackage[bookmarksnumbered,bookmarksopen,colorlinks,citecolor=blue,linkcolor=blue] {hyperref}

\begin{document}
\title{Production of new neutron-rich heavy nuclei with $Z=56-80$ in the multinucleon transfer reactions of $^{136}$Xe+$^{198}$Pt}
\author{Cheng Li$\,^{1,2}$}\email{licheng@mail.bnu.edu.cn}
\author{Fan Zhang$\, ^{3}$}
\author{Xinxin Xu$\, ^{1,2}$}
\author{Jingjing Li$\, ^{1,2}$}
\author{Gen Zhang$\, ^{1,2}$}
\author{Bing Li$\, ^{1,2}$}
\author{Cheikh A.T. Sokhna$\, ^{1,2}$}
\author{Zhishuai Ge$\, ^{1,2}$}
\author{Peiwei Wen$\, ^{4}$}
\author{Feng-Shou Zhang$\, ^{1,2,5}$}\email{Corresponding author:
fszhang@bnu.edu.cn}

\affiliation{
$^{1}$Beijing Radiation Center, Beijing 100875, China\\
$^{2}$College of Nuclear Science and Technology,
Beijing Normal University, Beijing 100875, China\\
$^{3}$Department of Electronic Information and Physics,
Changzhi University, Changzhi 046011, China\\
$^{4}$China Institute of Atomic Energy, Beijing 102413, China\\
$^{5}$Center of Theoretical Nuclear Physics, National Laboratory of Heavy Ion
Accelerator of Lanzhou, Lanzhou 730000, China
}


\begin{abstract}
The multinucleon transfer reactions in collisions of $^{136}$Xe+$^{198}$Pt at incident energies $E_{\textrm{lab}}=$5.25, 6.20, 7.98, 10.0, and 15.0 MeV/nucleon are investigated by using the improved quantum molecular dynamics model. It is found that 6.20 MeV/nucleon is the optimal incident energy for producing the neutron-rich heavy nuclei. About 80 unknown neutron-rich nuclei might be produced in this reaction with cross sections from 10$^{-6}$ to 10$^{-2}$ mb. The angular distributions of the neutron-rich isotopes are predicted.

\end{abstract}


\maketitle

\section{Introduction \label{intro}}

Producing new nuclei towards the neutron and proton drip lines in the nuclear landscape is a fundamental task in nuclear physics. At the end of 2017, a total of 3252 stable and radioactive nuclides have been discovered \cite{iso1}. In the proton-rich side of the nuclear chart, the known proton-rich nuclides are very close to the proton drip line up to $Z=91$. However, the neutron drip line nuclides are synthesized only up to $Z=8$ \cite{iso2}. A large number of neutron-rich nuclides are unknown due to the long distance separating the valley of stability from the neutron drip line. Many experiments were performed to produce the new neutron-rich nuclei via in-flight fission and cold-fragmentation of $^{238}$U at intermediate or relativistic energies. The in-flight fission reactions have proven to be an optimum method for the production of new medium-mass neutron-rich nuclei \cite{frag1,frag2,frag3}. For example, 36 new neutron-rich nuclei in the range of $37\leqslant Z\leqslant57$ were observed by using in-flight fission of a 345 MeV/nulceon $^{238}$U beam impinged on $^{9}$Be target at RIKEN \cite{frag3}. Another experiment of 1 GeV/nucloen $^{238}$U+$^{9}$Be performed at GSI shows that the new neutron-rich heavy nuclei range from $72\leqslant Z\leqslant78$ were produced by the cold-fragmentation method \cite{frag4}. However, production cross sections of these new nuclei from the cold-fragmentation mechanism are only a few ten pb orders of magnitude.

In recent years, the multinucleon transfer (MNT) reactions have attracted extensive attention to produce new neutron-rich nuclei both theoretically \cite{rew1,rew2,rew3,rew4,rew5,rew6,rew7,rew8,rew9,rew10,rew11,rew12,rew13,rew14,rew15,rew16,rew17} and experimentally \cite{exp1,exp2,exp3,exp4,exp5,exp6,exp7,exp8,exp9}. An experiment of $^{136}$Xe+$^{198}$Pt at incident energies $E_{lab}=7.98$ MeV/nucleon was performed by watanabe \emph{et al.} at GANIL \cite{exp3}. They found that the production cross sections of the $N=126$ isotones are extremely significant larger than those measured in the fragmentation reaction of a 1 GeV/nucleon $^{208}$Pb beam impinged on a Be target \cite{pb}. However, the new neutron-rich nuclei around $N=126$ shell closure were not observed because the detection limit of the VAMOS++ is 10$^{-2}$ mb. Thereafter, Zhu \emph{et al.} want to repeat this experiment using the Fragment Mass Analyzer (FMA) detector \cite{FMA} at Argonne National Laboratory. Another group, Huang \emph{et al.}, also have a plan to perform the $^{136}$Xe+$^{198}$Pt reaction with the gas-filled separator at High Intensity heavy-ion Accelerator Facility (HIAF) \cite{HIAF1,HIAF2}. Theoretical support for these very time-consuming and expensive experiments is vital for choosing the optimum incident energy and arranging the detector.

The models for describing the MNT reactions include the improved quantum molecular dynamics (ImQMD) model \cite{rew2,rew3,QMD0}, the time-dependent Hartree-Fock (TDHF) approach \cite{TDHF1,TDHF2}, the dinuclear system (DNS) model \cite{DNS1,DNS2,DNS3,DNS4}, the GRAZING model \cite{GRA1,GRA2,GRA3,GRA4}, the Complex WKB (CWKB) model \cite{CWKB1,CWKB2,CWKB3}, the Langevin equations \cite{rew1}, the deep-inelastic transfer (DIT) model \cite{DIT1,DIT2}, and so on.

In our previous work \cite{rew2}, we have studied the MNT reaction of $^{136}$Xe+$^{198}$Pt at 7.98 MeV/nucleon by using the ImQMD model. By analyzing the differential cross sections of the products, we have found that the new neutron-rich nuclei are produced in deep-inelastic collision mechanism. In this work, we will systematically investigate the $^{136}$Xe+$^{198}$Pt collisions at different incident energies. On the one hand, we will search for the optimal incident energy for the production of new neutron-rich nuclei. On the other hand, the emission angle of these new neutron-rich nuclei will be predicted for the experiment to arrange the detector.

The structure of this paper is as follows. In Sec. II, we briefly introduce the ImQMD model. The results and discussion are presented in Sec. III. Finally the conclusion is given in Sec. IV.

\section{Theoretical Framework}
The ImQMD model is an improved version of the quantum molecular dynamics (QMD) model \cite{QMD} which takes into account the effects of the surface term, the surface-symmetry potential term, and so on \cite{Co1,Co2}. To describe the fermionic nature of the $N$-body system and improve the stability of an individual nucleus, the Fermi constraint is adopted. The two-body collision correlations and the Pauli blocking checking are also included. The same as the original QMD model, each nucleon is represented by a coherent state of a Gaussian wave packet
\begin{equation}  \label{1}
\phi _{i}(\mathbf{r})=\frac{1}{(2\pi \sigma _{r}^{2})^{3/4}}\exp [-\frac{(
\mathbf{r-r}_{i})^{2}}{4\sigma
_{r}^{2}}+\frac{i}{\hbar}\mathbf{r}\cdot \mathbf{p}_{i}],
\end{equation}
where $\mathbf{r}_{i}$ and $\mathbf{p}_{i}$ are the centers of $i$th wave
packet in the coordinate and momentum space, respectively. $\sigma
_{r}$ represents the spatial spread of the wave packet in coordinate space. The one-body
phase space distribution function is obtained by the Wigner
transform of the wave function. The time evolution of $\mathbf{r}_{i}$ and $\mathbf{p}_{i}$ for each nucleon is governed by Hamiltonian equations of motion
\begin{equation}  \label{8}
\dot{\mathbf{r}}_{i}=\frac{\partial H}{\partial \mathbf{p}_{i}}, \dot{%
\mathbf{p}}_{i}=-\frac{\partial H}{\partial \mathbf{r}_{i}}.
\end{equation}
The Hamiltonian of the system includes the kinetic energy $T=\sum\limits_{i} \frac{\mathbf{p}_{i}^{2}}{2m}$ and effective
interaction potential energy
\begin{equation}  \label{9}
H=T+U_{\mathbf{Coul}}+U_{\mathbf{loc}},
\end{equation}
where, $U_{\mathbf{Coul}}$ is the Coulomb energy, which is written as a sum of the direct and
the exchange contribution
\begin{eqnarray}
 \nonumber U_{\mathbf{Coul}}=&\frac{1}{2}\int\int{\rho_{p}(\textbf{r})}
 \frac{e^{2}}{|\textbf{r}-\textbf{r}'|}{\rho_{p}(\textbf{r}')}d\textbf{r}d\textbf{r}'\\
 &-e^{2}\frac{3}{4}(\frac{3}{\pi})^{1/3}\int\rho_{p}^{4/3}d\textbf{r}.\label{16}
\end{eqnarray}
$\rho_{p}$ is the density distribution of protons of the system. The nuclear interaction potential energy $U_{\textrm{loc}}$ is obtained from the integration of the Skyrme energy density functional $U=\int V_{\textrm{loc}}(\mathbf{r})d\mathbf{r}$ without the spin-orbit term, which reads
\begin{eqnarray}
\nonumber V_{\mathbf{loc}}=&&\frac{\alpha }{2}\frac{\rho ^{2}}{\rho _{0}}+\frac{\beta }{\gamma +1}%
\frac{\rho ^{\gamma +1}}{\rho _{0}^{\gamma }}+\frac{\textsl{g}_{sur}}{2\rho _{0}}%
(\nabla \rho )^{2}\\
&&\ +\frac{C_{s}}{2\rho _{0}}(\rho ^{2}-\kappa _{s}(\nabla \rho
)^{2})\delta ^{2} + g_{\tau}\frac{\rho ^{\eta +1}}{\rho_{0}^{\eta
}}. \label{12}
\end{eqnarray}
Here $\rho=\rho_{n}+\rho_{p}$ is the nucleons density. $\delta=(\rho_{n}-\rho_{p})/(\rho_{n}+\rho_{p})$ is the isospin asymmetry. The parameter set IQ2 (see Table 1) adopted in this work are
the same as in the Refs. \cite{rew2,rew3}.

\begin{table}
\tabcolsep=1pt \caption{The model parameters (IQ2) adopted in this work.}
{\begin{tabular}{@{}cccccccccc@{}}

\hline\hline
   $\alpha$ & $\beta$ & $\gamma$ & $\textsl{g}_{sur}$ & $\textsl{g}_{\tau}$ & $\eta$ & $C_{S}$ & $\kappa_{s}$ & $\rho_{0}$ \\
  $(\textrm{MeV})$ & $(\textrm{MeV})$ & $$ & $(\textrm{MeV}\cdot \textrm{fm}^{2})$ & $(\textrm{MeV})$ & $$ & $(\textrm{MeV})$ & $(\textrm{fm}^{2})$ & $(\textrm{fm}^{-3})$ \\
\hline
   $-356$ & $303$ & $7/6$ & $7.0$ & $12.5$ & $2/3$ & $32.0$ & $0.08$ & $0.165$\\
\hline\hline
\end{tabular}}

\end{table}

In this work, we set $z$-axis as the beam direction and $x$-axis as the impact parameter direction. The initial distance of the center of mass between the projectile and target is 30 fm. We base on the parameter sets of IQ2 and set the wave-packet width $\sigma_r=1.3$ fm. The IQ2 has been tested for describing the fusion reactions \cite{QMD1,QMD2}, MNT reactions \cite{rew2,rew3,QMD0}, and fragmentation reactions \cite{QMD4,QMD5,QMD6}. The dynamic simulation is stopped at 2000 fm/$c$. And then the GEMINI code \cite{GEMI1,GEMI2} is used to deal with the subsequent de-excitation process.
The nuclear level densities in the GEMINI code are taken as a Fermi-gas form with default parameters.

\section{RESULTS AND DISCUSSION}
In this work, we just consider the case of two fragments after the collisions. We first check the ImQMD model for describing the isotopic distributions in MNT reactions. Fig. 1 shows the isotopic distributions for $Z=$52, 54, 56, 76, and 80 in $^{136}$Xe+$^{198}$Pt reaction at $E_{\textrm{lab}}=7.98$ MeV/nucleon. The thick folding lines and thick solid lines denote the results of ImQMD+GEMINI and GRAZING with including the evaporation, respectively. The thin folding lines and thin solid lines denote the primary fragment distributions of the ImQMD and GRAZING, respectively. The experiment data are taken from the Ref. \cite{exp3} which are denoted by open circles. It should be pointed out that the calculated fragment yield distributions from the ImQMD are filtered with an angular range $24^{\circ}-34^{\circ}$ (the grazing angle is about $33^{\circ}$). This range is the same as the experiment. The GRAZING calculations include only a small range of impact parameters close to the grazing where most of the cross section is concentrated in the quasielastic scattering. The contributions coming from small impact parameters, leading to deep inelastic collision, are not considered \cite{GRA1,GRA2}. From Fig. 1, one can see that the particle evaporation is extremely significant in the de-excitation processes both for the ImQMD and GRAZING model. The GRAZING model grossly underestimates the production cross section by orders of magnitude in the case of $Z=54$ and 76. The ImQMD+GEMINI calculations are reasonably well to describe the isotopic production cross section both for the projectile-like fragments (PLFs) and the target-like fragments (TLFs).

\begin{figure*}
\begin{center}
\includegraphics[width=16cm,angle=0]{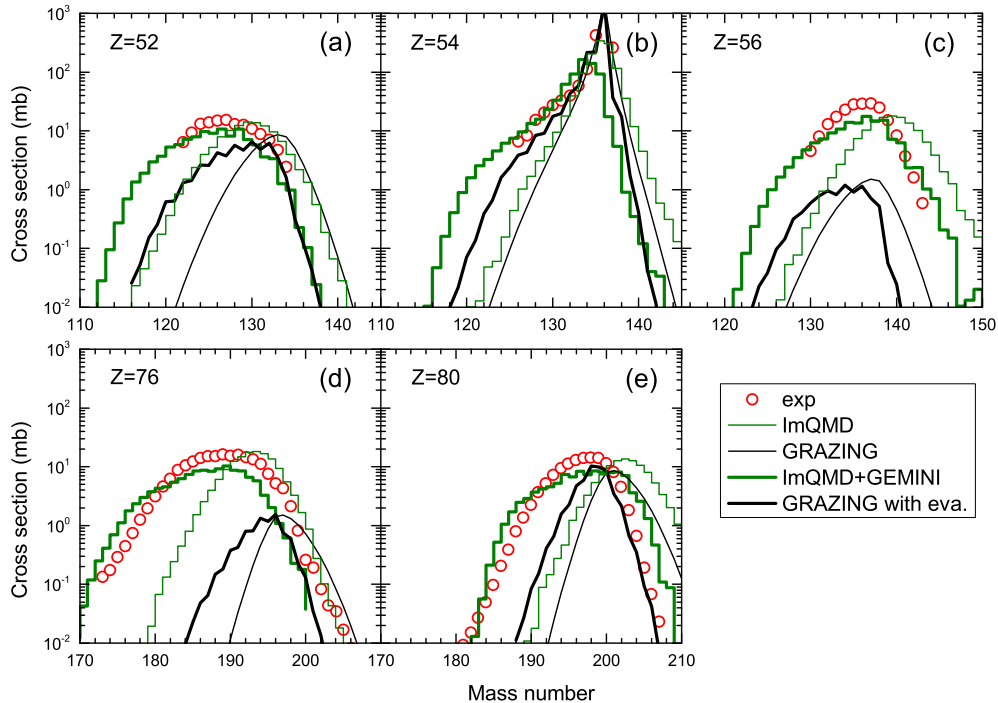}

\caption{
Isotopic distributions for $Z=$52, 54, 56, 76, and 80 in $^{136}$Xe+$^{198}$Pt reaction at $E_{\textrm{lab}}=7.98$ MeV/nucleon. The thick lines denote the secondary fragment distributions from the ImQMD+GEMINI and GRAZING with including evaporation. The thin lines are primary fragment distributions. The open circles denote the experiment data from the Ref. \cite{exp3}. }

\end{center}
\end{figure*}

In order to obtain the optimal incident energy for producing the neutron-rich heavy nuclei, we calculate systematically the $^{136}$Xe+$^{198}$Pt reactions at different incident energies. Fig. 2 shows the isotopic distributions of secondary fragments in $^{136}$Xe+$^{198}$Pt at $E_{\textrm{lab}}=$5.25, 6.20, 7.98, 10.0, and 15.0 MeV/nucleon. The range of the emission angle of the products is from 0$^\circ$ to 180$^\circ$. Two typical features can be found in the isotopic distributions at $Z=78$. The high peaks located at $A=198$ are mainly came from the contribution of quasielastic collisions where the primary products have low excitation energies. The bell shaped distributions in the neutron-deficient side correspond to more damped events with strong particle evaporation. Especially at large incident energies, a large number of particles are evaporated, which causes a shift of the final products to the neutron-deficient side. In order to obtain the neutron-rich nuclei, it is expected that the primary products have a large production probability and have a small excitation energy. From Fig. 2, we find that the optimal incident for producing the neutron-rich nuclei is 6.20 MeV/nucleon. Because for the case of $E_{\textrm{lab}}=$5.25 MeV/nucleon, the production cross sections of the primary neutron-rich nuclei are lower. While for a larger incident energy, the primary products are highly excited and thus leads to the lower survival probability for the neutron-rich nuclei. Especially in the proton pickup channels, the production cross sections of the neutron-rich nuclei are quite low due to a dominant fission channel.

In Fig. 3, we display the energy dissipation processes in MNT reactions. The excitation energy of an excited fragment is obtained as the total energy of the fragment in the body frame with the corresponding ground state binding energy being subtracted. Fig. 3(a) shows the average excitation energy as a function of the mass number of the TLFs in $^{136}$Xe+$^{198}$Pt at different incident energies. One can see that the energy dissipation of the system is strongly associated with the incident energy. A large incident energy improves significantly the excitation energy of the products, especially in the region of $A>$198. Hence, we can see the production cross sections of the neutron-rich nuclei decrease rapidly with increasing incident energy at the proton pickup channels (see Fig.2(d)-2(e)). Fig. 3(b) shows the average excitation energy as a function of the neutron number of the products in $^{136}$Xe+$^{198}$Pt at $E_{\textrm{lab}}=$ 6.20 MeV/nucleon. From Fig. 3(b), one sees that the average excitation energy decrease with increasing neutron number of the products. It is due to that the $Q_{gg}$ value (most are negative) is rapidly decreased with the increase of neutron number of the products. It is particularly important for the survival of the exotic neutron-rich nuclei. In Ref. \cite{exp3}, it is found that the very neutron-rich nuclei are dominantly produced in the collisions with the low total kinetic energy lost (TKEL).



\begin{figure*}
\begin{center}
\includegraphics[width=16cm,angle=0]{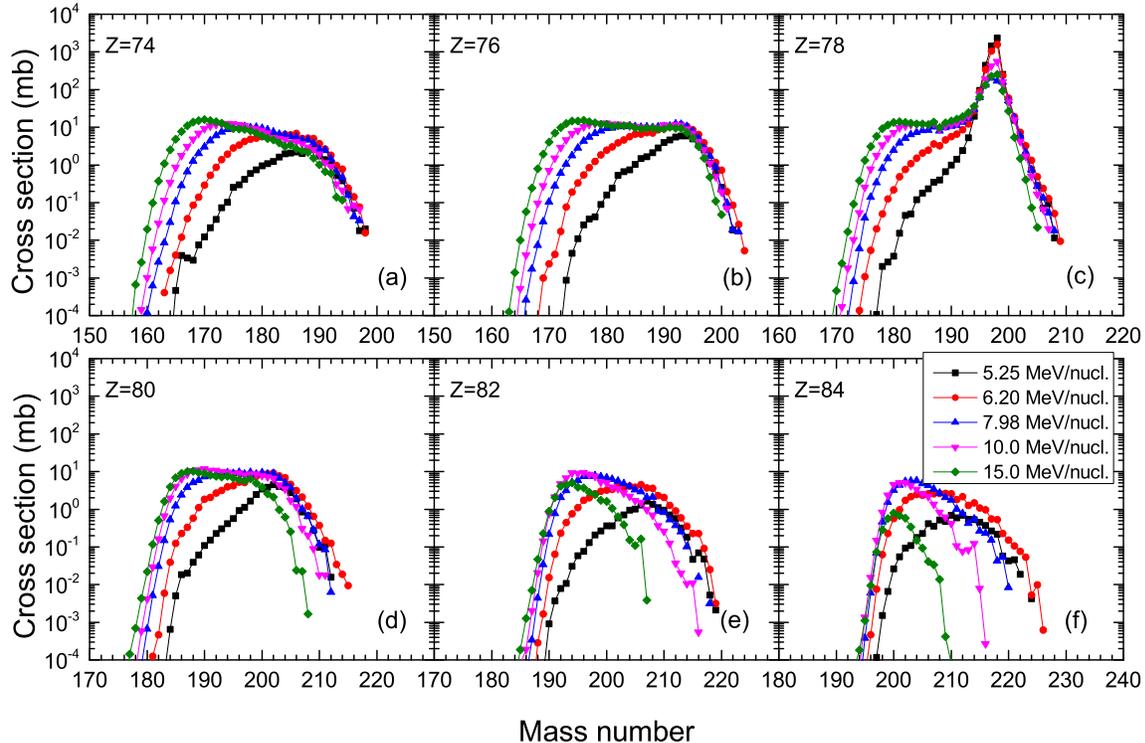}
\caption{
 Calculated isotopic distributions of secondary fragments by the ImQMD model in $^{136}$Xe+$^{198}$Pt at different incident energies.
}

\end{center}
\end{figure*}

\begin{figure*}
\begin{center}
\includegraphics[width=14cm,angle=0]{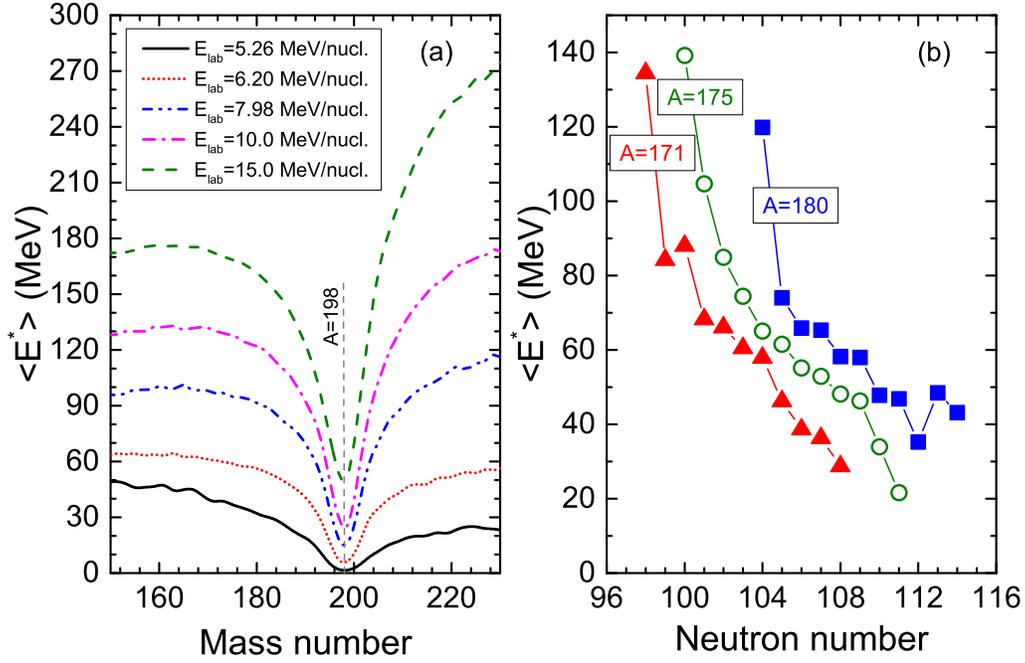}
\caption{
 (a) Average excitation energy as a function of the mass number of the TLFs in $^{136}$Xe+$^{198}$Pt at $E_{\textrm{lab}}=$5.26, 6.20, 7.98, 10.0, and 15.0 MeV/nucleon. (b) Average excitation energy as a function of the neutron number of the primary products in $^{136}$Xe+$^{198}$Pt at $E_{\textrm{lab}}=$ 6.20 MeV/nucleon.
}

\end{center}
\end{figure*}

Fig. 4 shows that the final production cross section distributions for TLFs in $^{136}$Xe+$^{198}$Pt at $E_{\textrm{lab}}=$ 6.20 MeV/nucleon. The folding lines denote the boundary of the known nuclei. From Fig. 4, one can see that about 80 new neutron-rich nuclei with atomic number from 56 to 80 are survived after the deexcitation processes. The production cross sections of these nuclei are from 10$^{-6}$-10$^{-2}$ mb. The new nuclei along the $N=126$ shell closure, $^{200}$W and $^{199}$Ta, could be produced in this reaction with cross sections 4.4 and 8.1 $\mu$b, respectively. In our previous work \cite{rew2}, it is indicated that the very neutron-rich isotopes are produced in deep-inelastic collisions. The exotic neutron-rich nuclei cannot be generated in quasielastic and quasifission collisions. Because in the quasielastic collisions, only a few nucleons could be transferred. While for the quasifission collisions, large excitation energy is obtained in energy dissipation processes which causes that the primary fragments have a smaller survival probability in the deexcitation processes.

\begin{figure*}
\begin{center}
\includegraphics[width=14cm,angle=0]{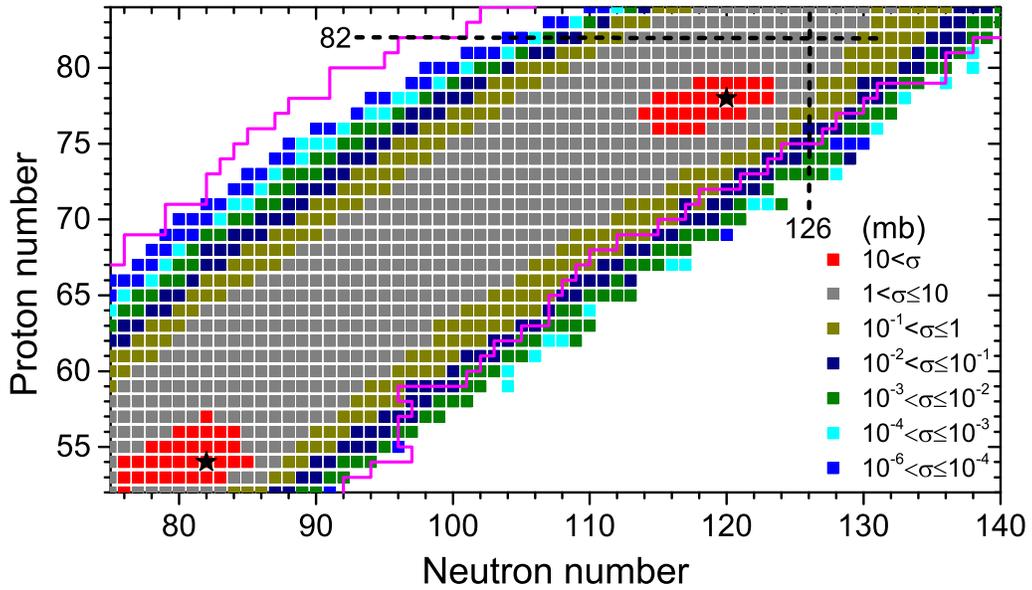}
\caption{
 Predicted chart of the products in $^{136}$Xe+$^{198}$Pt at $E_{\textrm{lab}}=6.20$ MeV/nucleon. The folding lines denote the boundary of the known nuclei. The stars denote the positions of the projectile and target.
}

\end{center}
\end{figure*}

\begin{figure*}
\begin{center}
\includegraphics[width=18cm,angle=0]{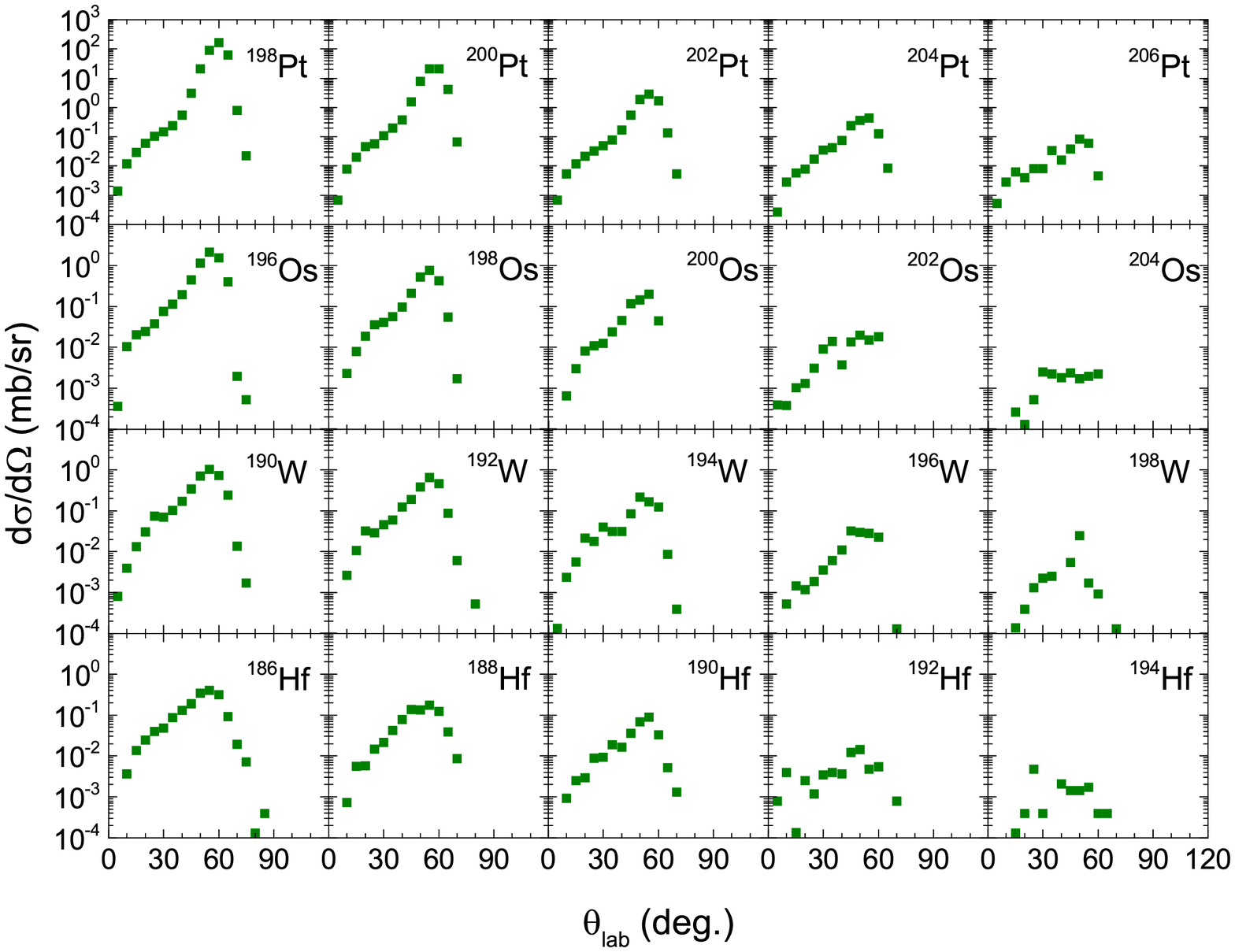}
\caption{
Predicted angular distributions of the TLFs in $^{136}$Xe+$^{198}$Pt at $E_{\textrm{lab}}=6.20$ MeV/nucleon.
}

\end{center}
\end{figure*}

Fig. 5 shows the angular distribution of the TLFs in $^{136}$Xe+$^{198}$Pt at $E_{\textrm{lab}}=6.20$ MeV/nucleon. The emission angle of the target in the grazing collision is about 60$^{\circ}$. From Fig. 5, one can see that the nuclei near the target (such as $^{190}$W, $^{196}$Os, $^{200}$Pt, etc.) could be produced by two reaction mechanisms. The products with emission angles around 60$^{\circ}$ come mainly from the contribution of the quasielastic collisions. The products with emission angles less than 40$^{\circ}$ are generated from the more damped collisions. With increasing of neutron excess, the emission angles of the products show a decreased trend. The exotic neutron-rich nuclei (such as  $^{206}$Pt, $^{204}$Os, $^{198}$W, and $^{194}$Hf) are mainly emitted from angles $\theta_{\textrm{lab}}<60^{\circ}$. Considering the relatively large production cross sections, $20^{\circ}<\theta_{\textrm{lab}}<60^{\circ}$ might be a suitable angular range to detect the new neutron-rich heavy nuclei.

\section{CONCLUSIONS}
The multinucleon transfer reactions of $^{136}$Xe+$^{198}$Pt at different incident energies have been studied by the ImQMD model. The results show that at low incident energies, the production cross sections of the primary neutron-rich nuclei are lower. While for a large incident energy, the primary neutron-rich products will obtain a high excitation energy which causes the lower survival probability in the de-excitation processes. We find that incident energy of 6.20 MeV/nucleon is optimal to produce the exotic neutron-rich heavy nuclei. The average excitation energy of the TLFs at different incident energies also have been analyzed. It shows that the energy dissipation of the system is strongly associated with the incident energy. The excitation energy of the products increase rapidly with increased mass transfer. In addition, we also find that the excitation energy decreases with increased neutron excess of the products. This is extremely advantageous for the survival of the neutron-rich nuclei. The angular distributions of the TLFs have been displayed. It shows that $20^{\circ}<\theta_{\textrm{lab}}<60^{\circ}$ might be a suitable angular range to detect the new neutron-rich heavy nuclei.

\section*{ACKNOWLEDGEMENTS}
This work was supported by the National Natural Science Foundation of China under Grants No. 11805015, No. 11805280, No. 11635003, No. 11025524, No. 11161130520, and No. 11605270;
the National Basic Research Program of China under Grant No. 2010CB832903; the European Commission's 7th Framework Programme (Fp7-PEOPLE-2010-IRSES) under Grant Agreement Project
No. 269131; the Project funded by China Postdoctoral Science Foundation (Grant No. 2016M600956, No. 2018T110069, and No. 2017M621035); the Beijing Postdoctoral Research Foundation
(2017-zz-076).

%

\end{document}